\documentclass{elsart}

\usepackage{amssymb}
\usepackage{amsmath}

\begin{document}

\begin{frontmatter}

\title{Nonlocal electrodynamics of accelerated systems}

\author{Bahram Mashhoon}

\address{Department of Physics and Astronomy\\
University of Missouri-Columbia\\
Columbia, MO 65211, USA}

\ead{mashhoonb@missouri.edu}

\begin{abstract}
Acceleration-induced nonlocality is discussed and a simple field theory
of nonlocal electrodynamics is developed. The theory involves a pair
of real parameters that are to be determined from observation. The
implications of this theory for the phenomenon of helicity-rotation
coupling are briefly examined.
\end{abstract}
\end{frontmatter}

\section{Introduction\label{sec:1}}

Consider the measurement of a basic \textit{radiation} field $\psi$ by
an accelerated observer in Minkowski spacetime. According to the hypothesis
of locality \cite{1}, the observer, at each event along its worldline, is locally equivalent to an otherwise identical momentarily comoving inertial observer. The frame of this hypothetical inertial observer is related to the background global inertial frame via a Poincar\'e transformation; therefore, the field measured by the momentarily comoving observer is $\widehat{\psi}(\tau )=\Lambda (\tau )\psi (\tau)$, where $\tau$ is the observer's proper time at the event under consideration and $\Lambda(\tau)$ is a matrix representation of the Lorentz group.

Let $\widehat{\Psi}$ be the field that is actually measured by the accelerated observer. The hypothesis of locality requires that $\widehat{\Psi} (\tau )=\widehat{\psi}(\tau ).$ However, the most general linear relation between $\widehat{\Psi}(\tau )$ and $\widehat{\psi}(\tau)$ consistent with causality is \cite{2}
\begin{equation}\label{eq:1}
\widehat{\Psi}(\tau )=\widehat{\psi}(\tau )+\int^\tau _{\tau _0} K(\tau ,\tau ')\widehat{\psi} (\tau ')d\tau ',
\end{equation}
where $\tau_0$ is the initial instant at which the observer's acceleration is turned on. The manifestly Lorentz-invariant ansatz \eqref{eq:1} involves a kernel that must be proportional to the acceleration of the observer. The kernel is determined from the postulate that a basic \textit{radiation} field can never stand completely still with respect to an accelerated observer. This is simply a generalization of the standard result for inertial observers. A detailed analysis reveals that the only physically acceptable kernel consistent with this physical requirement is \cite{3}-\cite{6}
\begin{equation}\label{eq:2}
K(\tau ,\tau  ')=k(\tau ')=-\frac{d\Lambda (\tau ')}{d\tau '} \Lambda^{-1}(\tau ').
\end{equation}
Using this kernel, Eq.~\eqref{eq:1} may be written as 
\begin{equation}\label{eq:3}
\widehat{\Psi}(\tau )=\widehat{\psi} (\tau _0)-\int^\tau _{\tau_0} \Lambda (\tau ')\frac{d\psi(\tau ')}{d\tau '}d\tau '.
\end{equation}
An immediate consequence of this relation is that if the accelerated observer passes through a spacetime region where the field $\psi$ is constant, then the accelerated observer measures a constant field as well, since $\widehat{\Psi}(\tau )=\hat{\psi}(\tau _0)$. This  is the main property of kernel \eqref{eq:2} and it will be used in the following section to argue that in nonlocal electrodynamics, Eq.~\eqref{eq:2} is only appropriate for the electromagnetic potential.

The basic notions that underlie this nonlocal theory of accelerated observers appear to be consistent with the quantum theory~\cite{7}-\cite{9}. Indeed, such an agreement has been the main goal of the nonlocal extension of the standard relativity theory of accelerated systems~\cite{10,11}. Moreover, the observational consequences of the theory are consistent with experimental data available at present. On the other hand, our treatment of nonlocal electrodynamics has thus far emphasized only \textit{radiation} fields. However, a nonlocal field theory of electrodynamics must also deal with special situations such as electrostatics and magnetostatics. Furthermore, the application of our nonlocal theory to electrodynamics encounters an essential ambiguity: should the basic field $\psi$ be identified with the vector potential $A_\mu$ or the Faraday tensor $F_{\mu\nu}$? In our previous treatments~\cite{10,12}, this ambiguity was left unresolved, since for the issues at hand either approach seemed to work. Nevertheless our measurement-theoretic approach to acceleration-induced nonlocality could be more clearly stated in terms of the directly measurable and gauge-invariant Faraday tensor, which was therefore preferred~\cite{10,12}.

The main purpose of the present work is to resolve this basic ambiguity in favor of the vector potential. The physical reasons for this choice are discussed in the following section. Section~\ref{sec:3} is then devoted to the determination of the appropriate kernel for the nonlocal Faraday tensor. Section~\ref{sec:4} deals with the consequences of this approach for the phenomenon of spin-rotation coupling for photons. The results are briefly discussed in section~\ref{sec:5}.

\section{Resolution of the ambiguity\label{sec:2}}

It is a consequence of the hypothesis of locality that an accelerated observer carries an orthonormal tetrad $\lambda^\mu_{\;\;(\alpha )}$. The manner in which this local frame is transported along the worldline reveals the acceleration of the observer; that is,
\begin{equation}\label{eq:4}
\frac{d\lambda^\mu _{\;\;(\alpha)}}{d\tau} =\phi_\alpha^{\;\;\beta} \lambda^\mu _{\;\;(\beta)},
\end{equation}
where $\phi_{\alpha \beta}=-\phi_{\beta\alpha} $ is the antisymmetric acceleration tensor.

Let us now consider the determination of an electromagnetic field, with vector potential $A_\mu$ and Faraday tensor $F_{\mu \nu}$,
\begin{equation}\label{eq:5}
F_{\mu\nu}=\partial_\mu A_\nu -\partial _\nu A_\mu,
\end{equation}
by the accelerated observer. The measurements of the momentarily comoving inertial observers along the worldline are given by
\begin{equation}\label{eq:6}
\widehat{A}_\alpha =A_\mu \lambda^\mu_{\;\;(\alpha)},\quad \widehat{F}_{\alpha \beta} =F_{\mu\nu}\lambda^\mu_{\;\;(\alpha)}\lambda^\nu_{\;\;(\beta)}.
\end{equation}
Thus according to our basic ansatz~\cite{2}, the fields as measured by the accelerated observer are
\begin{align}\label{eq:7}
\widehat{\mathcal{A}}_\alpha (\tau )&=\widehat{A}_\alpha (\tau )+\int^\tau_{\tau _0}K_\alpha^{\;\;\beta}(\tau ,\tau ')\widehat{A}_\beta (\tau')d\tau',\\
\label{eq:8}\widehat{\mathcal{F}}_{\alpha \beta} (\tau )&=\widehat{F}_{\alpha \beta} (\tau )+\int^\tau _{\tau_0} K_{\alpha \beta} ^{\;\;\;\;\gamma \delta} (\tau ,\tau ')\widehat{F}_{\gamma\delta }(\tau')d\tau '.\end{align}
Though these relations may be reminiscent of the phenomenological memory-dependent electrodynamics of certain continuous media~\cite{13}, they do in fact represent field determinations in vacuum and are consistent---in the case of kernels~\eqref{eq:9} and \eqref{eq:11} specified below---with the averaging viewpoint developed by Bohr and Rosenfeld~\cite{14}.

It remains to determine the kernels in Eqs.~\eqref{eq:7} and \eqref{eq:8}. Specifically, which one should be identified with the result given in Eq.~\eqref{eq:2}? The aim of the following considerations is the construction of the simplest tenable nonlocal electrodynamics; however, there is a lack of definitive experimental results that could guide such a development. We must therefore bear in mind the possibility that future experimental data may require a revision of the theory presented in this paper.

Let us recall here the main property of kernel~\eqref{eq:2} noted in the previous section: a uniformly moving observer enters a region of constant field $\psi$; the observer is then accelerated, but it continues to measure the same constant field. Now imagine such an observer in an extended region of constant electric and magnetic fields; we intuitively expect that as the velocity of the observer varies, the electromagnetic field measured by the observer would in general vary as well. This expectation appears to be provisionally consistent with the result of Kennard's experiment~\cite{15,16}. It follows that the kernel in Eq.~\eqref{eq:8} cannot be of the form given in Eq.~\eqref{eq:2}. On the other hand, in a region of constant vector potential $A_\mu$, the gauge-dependent potential measured by an arbitrary accelerated observer could be constant; in fact, in this region the gauge-invariant electromagnetic field vanishes for all observers by Eqs.~\eqref{eq:5}, \eqref{eq:6} and \eqref{eq:8}. Therefore, we assume that the kernel in Eq.~\eqref{eq:7} is of the form given by Eq.~\eqref{eq:2}, so that
\begin{equation}\label{eq:9}
K_\alpha^{\;\;\beta} (\tau ,\tau ')=k_\alpha^{\;\;\beta} (\tau '),\end{equation}
which can be expressed via Eqs.~\eqref{eq:2} and \eqref{eq:4} as
\begin{equation}\label{eq:10} k_\alpha^{\;\;\beta}=-\phi_\alpha^{\;\;\beta}.\end{equation}
The determination of the field kernel in Eq.~\eqref{eq:8} is the subject of the next section.

\section{Field kernel\label{sec:3}}

The first step in the determination of the kernel in Eq.~\eqref{eq:8} is to require that
\begin{equation}\label{eq:11}
K_{\alpha \beta}^{\;\;\;\;\gamma\delta}(\tau ,\tau ')=k_{\alpha \beta}^{\;\;\;\;\gamma \delta} (\tau ').\end{equation}
This simplifying assumption is rather advantageous~\cite{4}-\cite{6}. If the acceleration of the observer is turned off at $\tau =\tau _f$, then the new kernel vanishes for $\tau >\tau _f$. In this case, the nonlocal contribution to Eq.~\eqref{eq:8} is a constant memory  of the past acceleration of the observer that is in principle measurable. This constant memory is simply canceled in a measuring device whenever the device is reset.

Next, we assume that $k_{\alpha \beta}^{\;\;\;\;\gamma \delta}$ is linearly dependent upon the acceleration tensor $\phi_{\alpha \beta}$. Clearly, the basic notions of the nonlocal theory cannot a priori exclude terms in the kernel that would be nonlinear in the acceleration of the observer. Therefore, our linearity assumption must be regarded as preliminary and contingent upon agreement with observation.

We have argued in the previous section that the electromagnetic field kernel given by Eq.~\eqref{eq:2}, which turns out to be 
\begin{equation}\label{eq:12} \kappa_{\alpha \beta}^{\;\;\;\;\gamma \delta}=-\frac{1}{2} (\phi _\alpha^{\;\;\gamma} \delta _{\beta}^{\;\;\delta} +\phi _\beta^{\;\;\delta}\delta_\alpha^{\;\;\gamma} -\phi_\beta^{\;\;\gamma} \delta_\alpha^{\;\;\delta} -\phi_\alpha ^{\;\;\delta }\delta_\beta ^{\;\;\gamma }),\end{equation}
cannot be the correct kernel by itself. To proceed, we must employ the Minkow\-ski metric tensor $\eta_{\alpha \beta}$, the Levi-Civita tensor $\epsilon_{\alpha \beta \gamma \delta}$ (with $\epsilon_{0123}=1$) and terms linear in the acceleration tensor $\phi_{\alpha \beta}(\tau )$ to generate kernels of the form $\kappa_{\alpha \beta}^{\;\;\;\;\gamma \delta} (\tau)$ that are antisymmetric in their first and second pairs of indices. A detailed discussion of such ``constitutive'' tensors is contained in~\cite{6}. It appears that all such kernels are linear combinations of Eq.~\eqref{eq:12} and its duals. The left dual results in a kernel given by
\begin{equation}\label{eq:13} ^\ast\kappa_{\alpha \beta}^{\;\;\;\;\gamma \delta} =\frac{1}{2} \epsilon_{\alpha \beta}^{\;\;\;\;\rho \sigma} \kappa_{\rho \sigma}^{\;\;\;\;\gamma \delta}.\end{equation}
This turns out to be equal to the kernel formed from the right dual, namely,
\begin{equation}\label{eq:14} \frac{1}{2} \kappa_{\alpha \beta}^{\;\;\;\;\rho \sigma} \epsilon_{\rho \sigma}^{\;\;\;\;\gamma\delta}=-\frac{1}{2} (\phi_\alpha^{\;\;\rho} \epsilon_{\rho\beta}^{\;\;\;\;\gamma\delta} -\phi_\beta^{\;\;\rho }\epsilon_{\rho\alpha}^{\;\;\;\;\gamma \delta}).\end{equation}
The equality of right and left duals in this case is due to $\phi_{\alpha\beta}=-\phi_{\beta \alpha}$ and simply follows from a general identity given on p.~255 of Ref.~\cite{6}. In connection with the general discussion of the invariants of the constitutive tensor in~\cite{6}, let us observe that $\kappa^\gamma_{\;\;\alpha\gamma \beta} =-\phi_{\alpha\beta}$, so that $\kappa_{\alpha\beta}^{\;\;\;\;\alpha\beta}=0$ and 
\begin{equation}\label{eq:15} \frac{1}{2}\kappa_{\gamma\delta}^{\;\;\;\;\rho\sigma} \kappa_{\rho\sigma}^{\;\;\;\;\gamma\delta}=-\phi_{\alpha \beta}\phi^{\alpha\beta}.\end{equation}
 Finally, the mixed duals vanish; for instance,
\begin{equation}\label{eq:16} \frac{1}{2}\kappa_{\alpha \rho \sigma \beta} \; \epsilon^{\rho \sigma\gamma \delta}\end{equation}
results in a kernel of the form
\begin{equation}\label{eq:17} \zeta_{\alpha \beta}^{\;\;\;\;\gamma\delta}=\frac{1}{4}(\kappa_{\alpha \rho \sigma \beta}-\kappa_{\beta \rho \sigma \alpha})\epsilon^{\rho \sigma \gamma\delta},\end{equation}
which is identically zero due to the antisymmetric nature of $\phi_{\alpha \beta}$.

The above considerations suggest that a natural choice for kernel~\eqref{eq:11} would be
\begin{equation}\label{eq:18} k_{\alpha \beta}^{\;\;\;\;\gamma\delta} (\tau )=p\; \kappa _{\alpha\beta}^{\;\;\;\;\gamma\delta} (\tau )+q\; {^\ast\kappa}_{\alpha \beta}^{\;\;\;\;\gamma\delta} (\tau ),\end{equation}
where $p$ and $q$ are constant real numbers such that  $(p,q)\neq (1,0)$. These numerical coefficients may be determined from the comparison of the theory with observation. It is interesting to note that $\kappa_{\alpha \beta \gamma \delta}=-\kappa_{\gamma \delta \alpha \beta}$,
\begin{equation}\label{eq:19} {^\ast\kappa}_{\alpha\beta}^{\;\;\;\;\gamma\delta}=\frac{1}{2} (\epsilon_{\alpha\beta}^{\;\;\;\;\rho\gamma} \phi_\rho^{\;\;\delta}-\epsilon_{\alpha\beta}^{\;\;\;\;\rho\delta } \phi_\rho^{\;\;\gamma}),\end{equation}
and $\kappa$ is minus the right dual of ${^\ast\kappa}$, namely,
\begin{equation}\label{eq:20} \kappa_{\alpha \beta}^{\;\;\;\;\gamma\delta} =-\frac{1}{2}\;{^\ast\kappa}_{\alpha\beta}^{\;\;\;\;\rho\sigma} \epsilon_{\rho\sigma}^{\;\;\;\;\gamma\delta}.\end{equation}
The implications of the new field kernel \eqref{eq:18} for the phenomenon of helicity-rotation coupling may be explored with a view towards possibly limiting the range of $(p,q)$. This is done in the next section.

\section{Spin-rotation coupling\label{sec:4}}

Consider the measurement  of the electromagnetic field by observers that rotate uniformly with frequency $\Omega_0>0$ about the direction of propagation of an incident plane monochromatic electromagnetic wave of frequency $\omega >0$. Specifically, we imagine a global inertial frame with coordinates $(t,x,y,z)$ and a class of observers that move uniformly along straight lines parallel to the $y$ axis for $-\infty <t<0$, but at $t=0$ are forced to move on counterclockwise circular paths about the $z$ axis, which coincides with the direction of wave propagation. The signature of $\eta_{\alpha\beta}$ is assumed to be $+2$ and units are chosen such that $c=1$. For a typical observer with $z=z_0$, $x=r>0$ and $y=r\Omega_0t$ for $-\infty <t<0$ and for $t\geq 0$, $x=r\cos \varphi$ and $y=r\sin \varphi$, where $\varphi =\Omega_0t=\gamma \Omega_0\tau$. Here $\gamma$ is the Lorentz factor corresponding to $v=r\Omega_0$ and $\tau$ is the proper time of the observer. The natural tetrad frame of the observer in $(t,x,y,z)$ coordinates is given for $t\geq 0$ by
\begin{align}\label{eq:21} \lambda^\mu_{\;\;(0)} &=\gamma (1,-v\sin\varphi ,v\cos \varphi ,0),\\
\label{eq:22}\lambda^\mu_{\;\;(1)}&=(0,\cos \varphi ,\sin \varphi ,0),\\
\label{eq:23} \lambda^\mu _{\;\;(2)}&=\gamma (v,-\sin \varphi ,\cos \varphi ,0),\\
\label{eq:24}\lambda^\mu _{\;\;(3)}&=(0,0,0,1).\end{align}

The acceleration tensor $\phi_{\alpha\beta}$ in Eq.~\eqref{eq:4} can be decomposed as $\phi_{\alpha \beta}\mapsto (-\mathbf{g},\mathbf{\Omega})$ in analogy with the Faraday tensor. Here the ``electric" part $(\phi_{0i}=g_i)$ represents the translational acceleration of the observer, while the ``magnetic" part $(\phi_{ij}=\epsilon_{ijk}\Omega^k)$ represents the frequency of rotation of the observer's spatial frame with respect to a nonrotating (i.e., Fermi-Walker transported) frame. The scalar invariants $\mathbf{g}$ and $\mathbf{\Omega}$ completely characterize the acceleration of the observer.

A typical rotating observer under consideration here has a centripetal acceleration $\mathbf{g}=-v\gamma^2\Omega_0(1,0,0)$ and rotation frequency $\mathbf{\Omega} =\gamma^2\Omega_0 (0,0,1)$ with respect to the local spatial frame $\lambda^\mu _{\;\;(i)}$, $i=1,2,3$, that indicate the radial, tangential and $z$ directions, respectively.

In an incident plane monochromatic wave of positive (negative) helicity, the electric and magnetic fields rotate counterclockwise (clockwise) about the direction of wave propagation. The frequency of this rotation is equal to the wave frequency $\omega\; (-\omega)$. Now imagine, as in the previous paragraph, observers rotating about the direction of wave propagation with frequency $\Omega_0\ll\omega$. According to such observers, the electric and magnetic fields rotate with frequency $\omega-\Omega_0 \; (-\omega-\Omega_0)$ about the direction of wave propagation. Thus a typical observer perceives an incident wave of positive (negative) helicity with frequency $\widehat{\omega}=\gamma (\omega \mp\Omega_0)$, where the upper (lower) sign refers to a wave of positive (negative) helicity. Here $\gamma$ is the Lorentz factor of the observer and takes due account of time dilation. The intuitive account of helicity-rotation coupling presented here emerges from the simple kinematics of Maxwell's theory~\cite{17} and has a solid observational basis~\cite{17}-\cite{20}. In particular, it is responsible for the phenomenon of \textit{phase wrap-up} in the GPS system~\cite{18,19}.

An important aspect of helicity-rotation coupling for $\omega\gg\Omega_0$ that is crucial for choosing the correct field kernel is that the helicity of the wave and hence its state of polarization should be the same for both the rotating and the static inertial observers. Thus the nonlocal part of Eq.~\eqref{eq:8} should conform to this notion of chirality preservation. 

To study kernel~\eqref{eq:18} for the rotating observers under consideration here, it is useful to employ the decomposition $F_{\mu\nu} \mapsto (\mathbf{E},\mathbf{B})$ and replace $F_{\mu\nu}$ by a column $6$-vector $F$ that has $\mathbf{E}$ and $\mathbf{B}$ as its components, respectively. In this way, Eq.~\eqref{eq:8} can be regarded as a matrix equation such that the kernel is a $6\times 6$ matrix. The incident electromagnetic wave can then be represented as
\begin{equation}\label{eq:25} F_\pm (t,\mathbf{x})=i\omega A_{\pm} \begin{bmatrix} \mathbf{e}_\pm\\ \mathbf{b}_\pm\end{bmatrix} e^{-i\omega (t-z)}, \end{equation} where $A_\pm$ is a constant amplitude, $\mathbf{e_\pm}=(\widehat{\mathbf{x}} \pm i\widehat{\mathbf{y}} )/ \sqrt2 ,$ $\mathbf{b}_\pm =\mp i\mathbf{e}_\pm$ and the upper (lower) sign represents positive (negative) helicity radiation. The unit circular polarization vectors $\mathbf{e}_\pm$ are such that $\mathbf{e}_\pm \cdot \mathbf{e}^\ast_{\pm}=1$. Our basic ansatz~\eqref{eq:1} is linear; therefore, we use complex fields and adopt the convention that only their real parts are physically significant. 

Along the worldline of a rotating observer, the field measured by the momentarily comoving inertial observers is given by~\cite{8}
\begin{equation}\label{eq:26} \widehat{F}_\pm (\tau )=i\gamma \omega A_\pm \begin{bmatrix} \widehat{\mathbf{e}}_\pm \\ \widehat{\mathbf{b}}_\pm \end{bmatrix}e^{-i\widehat{\omega} \tau +i\omega z_0},\end{equation}
where $\widehat{\mathbf{b}}_\pm =\mp i\widehat{\mathbf{e}}_\pm $ and
\begin{equation}\label{eq:27} \widehat{\mathbf{e}}_\pm =\frac{1}{\sqrt2} \begin{bmatrix} 1\\ \pm i\gamma^{-1}\\ \pm iv \end{bmatrix} \end{equation}
are unit vectors with $\widehat{\mathbf{e}}_\pm \cdot \widehat{\mathbf{e}}_\pm ^\ast=1$. Here $\widehat{\omega} =\gamma (\omega \mp \Omega_0)$, which indicates the modification of the transverse Doppler effect by the helicity-rotation coupling. A significant implication of the hypothesis of locality is that by a mere rotation of frequency $\Omega_0=\omega$, the accelerated observer can stand completely still with respect to the incident positive-helicity radiation~\cite{8}. Another general consequence of the hypothesis of locality should also be noted: the relative amplitude of the helicity states $(A_+/ A_-)$ is not affected by the rotation of the observer~\cite{8}. It is important to examine how these conclusions are modified by the nonlocal theory presented here.

It follows from Eqs.~\eqref{eq:8}, \eqref{eq:10} and \eqref{eq:18} that the kernel in matrix notation is given by
\begin{equation}\label{eq:28} k=p\;\kappa +q\;{^\ast \kappa},\end{equation}
where
\begin{equation}\label{eq:29} \kappa =\begin{bmatrix} \kappa_1 & -\kappa_2\\ \kappa_2 &\kappa_1 \end{bmatrix} ,\quad {^\ast\kappa} =\begin{bmatrix} -\kappa_2 & -\kappa _1\\ \kappa_1 & -\kappa_2 \end{bmatrix} .\end{equation}
Here $\kappa_1 =\mathbf{\Omega} \cdot\mathbf{I}$ and $\kappa_2 =\mathbf{g}\cdot \mathbf{I}$, where $I_i$, $(I_i)_{jk}=-\epsilon _{ijk}$, is a $3\times3$ matrix proportional to the operator of infinitesimal rotations about the $x^i$ axis.

Using kernel~\eqref{eq:28}, we find that the field measured by the accelerated observer is
\begin{equation}\label{eq:30} \widehat{\mathcal{F}}_\pm (\tau )=\widehat{F}_\pm (\tau ) \left[ 1+\frac{(\pm p+iq)\Omega_0 }{\omega \mp \Omega_0} (1-e^{i\widehat{\omega}\tau })\right].\end{equation}
Note that $\widehat{\mathcal{F}} _\pm$ can become constant---that is, the incident wave can stand still with respect to the accelerated observer---for $\omega \mp \Omega_0 =-(\pm p+iq)\Omega_0$, which is impossible so long as $q\neq 0$. Henceforth we assume that $q$ does not vanish. For positive-helicity incident radiation at the resonance frequency $\omega =\Omega_0$,
\begin{equation}\label{eq:31} \widehat{\mathcal{F}}_+ (\tau )=\widehat{F}_+ [1-i(p+iq)\gamma \Omega _0\tau ],\end{equation}
where $\widehat{F}_+$ is constant. Thus the rotating observer does not stand still with the wave as a direct consequence of nonlocality; moreover, the linear divergence with time in Eq.~\eqref{eq:31} would disappear for a finite incident pulse of radiation. Next, Eq.~\eqref{eq:30} implies that the ratio of the measured amplitude of positive-helicity radiation to that of negative-helicity radiation is $(A_+/A_-)\rho$, where $\rho $ is given by
\begin{equation}\label{eq:32} \rho =\frac{\omega^2-\Omega_0^2+\Omega_0 (\omega +\Omega_0)(p+iq)}{\omega^2-\Omega_0^2-\Omega_0(\omega -\Omega_0)(p-iq)}.\end{equation}
It follows from previous results~\cite{8} that we should expect $|\rho |>1$ for $\omega^2>\Omega^2_0$; in fact, Eq.~\eqref{eq:32} implies that $|\rho |>1$ whenever
\begin{equation}\label{eq:33} p^2+q^2+p\left( \frac{\omega^2}{\Omega_0^2}-1\right)>0.\end{equation}
This relation is satisfied for $\omega^2 >\Omega^2_0$ when $p\geq 0$. These results should be compared and contrasted with similar ones given for $(p,q)=(1,0)$ in \cite{8}, where nonlocal electrodynamics is indirectly tested by comparing its consequences with the standard quantum mechanics of the interaction of photons with rotating electrons in the correspondence limit. One may conclude from our analysis of the spin-rotation coupling in this section that in kernel~\eqref{eq:18} $p$ and $q$ should be such that $p\geq 0$, $p\neq 1$ and $q\neq 0$. It is interesting to note that for $q\neq 0$, there is a certain nonlocality-induced helicity-acceleration coupling in the complex amplitude of the field measured by an observer that is linearly accelerated along the direction of incidence of a plane electromagnetic wave~\cite{7}. It seems that further restrictions on $p$ and $q$ should be based on observational data.

\section{Discussion\label{sec:5}}

A foundation has been laid for the simplest nonlocal field theory of electrodynamics appropriate for accelerated systems. The postulated determination of memory-dependent quantities in Eqs.~\eqref{eq:7} and \eqref{eq:8} may be interpreted in terms of the projection of certain nonlocal field variables on the local tetrads. That is, we can define $\mathcal{A}_\mu$ and $\mathcal{F}_{\mu\nu}$ via
\begin{equation}\label{eq:34} \widehat{\mathcal{A}}_\alpha =\mathcal{A}_\mu \lambda^\mu_{\;\;(\alpha)},\quad \widehat{\mathcal{F}}_{\alpha \beta} =\mathcal{F} _{\mu\nu}\lambda^\mu_{\;\; (\alpha)} \lambda^\nu_{\;\;(\beta)}.\end{equation}
Thus for a whole class of accelerated observers,  the resolvent kernels in Eqs.~\eqref{eq:7} and \eqref{eq:8} may be employed together with Eq.~\eqref{eq:34} to derive nonlocal field equations for $\mathcal{A}_\mu$ and $\mathcal{F}_{\mu\nu}$ as already illustrated in~\cite{12}. The resulting Maxwell equations for $\mathcal{F}_{\mu\nu}$ would then supersede the special source-free case with $(p,q)=(1,0)$ discussed in~\cite{12}. Moreover, Eq.~\eqref{eq:5} would lead to a complicated nonlocal relationship between $\mathcal{F}_{\mu\nu}$ and the gauge-dependent potential $\mathcal{A}_\mu$. A more complete discussion of these and related issues will be presented elsewhere.

\section*{Acknowledgements}

I am grateful to Friedrich Hehl for many valuable discussions. Thanks are also due to Yuri Obukhov for helpful correspondence.

\end{document}